\documentclass[superscriptaddress, reprint, amsmath, amssymb, aps, physrev, floatfix, nofootinbib, longbibliography]{revtex4-2}
\usepackage[english]{babel}
\usepackage[
    colorlinks=true,
    frenchlinks=false,
    linkcolor=blue,
    anchorcolor=blue,
    citecolor=blue,
    filecolor=blue,
    urlcolor=blue,
    bookmarks=true,
    bookmarksopen=true,
    bookmarksnumbered=true,
    bookmarksopenlevel=1,
    plainpages=false,
    pdfpagelabels=true,
    draft=false]{hyperref}
\usepackage{cleveref}
\usepackage{csquotes}
\usepackage{unicode-math}

\usepackage{graphicx}
\usepackage[separate-uncertainty=true, multi-part-units=single]{siunitx} 
\usepackage{color} 
\usepackage{braket}
\DeclareSIUnit\dBm{dBm}
\DeclareSIUnit\bar{bar}
\newcommand{\mode}{\hat{a}}

\usepackage[mathlines]{lineno} 
\begin{document}

\title{Quantum secret sharing in tripartite superconducting network}

% list of authors
\author{W.~K.~Yam}
\email{WunKwan.Yam@wmi.badw.de}
\affiliation{Walther-Mei{\ss}ner-Institut, Bayerische Akademie der Wissenschaften, 85748 Garching, Germany}
\affiliation{School of Natural Sciences, Technical University of Munich, 85748 Garching, Germany}

\author{C.~Wilkinson}
\affiliation{School of Physics and Astronomy, University of St. Andrews, United Kingdom}

\author{S.~Gandorfer}
\affiliation{Walther-Mei{\ss}ner-Institut, Bayerische Akademie der Wissenschaften, 85748 Garching, Germany}
\affiliation{School of Natural Sciences, Technical University of Munich, 85748 Garching, Germany}

\author{F.~Fesquet}
\affiliation{Walther-Mei{\ss}ner-Institut, Bayerische Akademie der Wissenschaften, 85748 Garching, Germany}
\affiliation{School of Natural Sciences, Technical University of Munich, 85748 Garching, Germany}

\author{M.~Handschuh}
\affiliation{Walther-Mei{\ss}ner-Institut, Bayerische Akademie der Wissenschaften, 85748 Garching, Germany}
\affiliation{School of Natural Sciences, Technical University of Munich, 85748 Garching, Germany}

\author{A.~Marx}
\affiliation{Walther-Mei{\ss}ner-Institut, Bayerische Akademie der Wissenschaften, 85748 Garching, Germany}

\author{R.~Gross}
\affiliation{Walther-Mei{\ss}ner-Institut, Bayerische Akademie der Wissenschaften, 85748 Garching, Germany}
\affiliation{School of Natural Sciences, Technical University of Munich, 85748 Garching, Germany}
\affiliation{Munich Center for Quantum Science and Technology (MCQST), 80799 Munich, Germany}

\author{N.~Korolkova}
\affiliation{School of Physics and Astronomy, University of St. Andrews, United Kingdom}

\author{K.~G.~Fedorov}
\email{Kirill.Fedorov@wmi.badw.de}
\affiliation{Walther-Mei{\ss}ner-Institut, Bayerische Akademie der Wissenschaften, 85748 Garching, Germany}
\affiliation{School of Natural Sciences, Technical University of Munich, 85748 Garching, Germany}
\affiliation{Munich Center for Quantum Science and Technology (MCQST), 80799 Munich, Germany}

\begin{abstract}
Superconducting microwave quantum networks is a rapidly developing field, enabling distributed quantum computing and holding a promise for hybrid architectures in quantum internet. Quantum secret sharing (QSS) is one of the key protocols for multipartite quantum networks and can provide an unconditionally secure way to share quantum states among $n$ players. Using microwave two-mode squeezed states as an entanglement resource, we experimentally implement a QSS protocol with $n = 3$, where a subset of at least $k = 2$ players must collaborate to faithfully reconstruct the original secret state. We demonstrate reconstructed-state fidelities that surpass the asymptotic no-cloning threshold of $F_\textrm{nc} = 2/3$ and identify a parameter regime that allows for unconditionally secure communication in the presence of an omnipotent dishonest player. Furthermore, we experimentally explore inherent connections between QSS and other important quantum information processing tasks, such as quantum dense coding and elementary quantum error correction of channel erasures. Finally, we discuss extensions of QSS and its relation to the concept of blind quantum computing.
\end{abstract}

\maketitle

Multipartite quantum networks possess many unique and useful features that enable novel secure communication protocols. The development of these protocols is essential to the implementation of a future quantum internet. Microwave quantum networks are inspired by the tremendous pace in the development of superconducting quantum circuits at GHz frequencies over the last two decades. One of the central benefits of such microwave quantum networks is their ability to advance quantum computing architectures from individual processor units to distributed computing systems, helping to solve the scaling challenge~\cite{Bravyi2022,Martinis2025}. Moreover, the emergence of open-air, microwave quantum communication presents a further exciting opportunity for network design~\cite{Casariego2023}. It has recently been estimated that continuous-variable (CV) quantum cryptographic protocols using microwaves can yield positive secret key rates over short distances in the open-air conditions~\cite{Pirandola2021,Fesquet2023}. Most importantly, it is suggested that microwave CV quantum key distribution is substantially more robust against open-air weather imperfections than its optical counterpart~\cite{Fesquet2023}.

Overall, we envision a concept of interconnected, large-scale hybrid quantum networks, which combine heterogeneous quantum information processing units and communication links, operating in both the microwave and optical regimes, both in-fiber and open-air. We demonstrate here the quantum secret sharing (QSS) protocol in the CV microwave regime, with the potential to serve as a fundamental building block for such hybrid quantum networks.

%\THz frequency regime is a recent emerging development in addition to the microwave and optical regimes -- KF: I removed this part for now, because I got slightly pessemistic about the THz regime recently - seemingly, it has not benefits of neither optics nor microwaves; it got the worst of the both worlds - but let's talk about it if necessary or if I'm missing something.

QSS offers an unconditionally secure way to distribute a quantum state between parties in a multi-node network. This secret quantum state is shared among $n$ players using quantum-entangled resource states. The original secret can only be faithfully reconstructed if at least $k > n/2$ players collaborate. Such a QSS protocol, where the secret is split into $n$ shares and at least $k$ are required to reconstruct it, is classed as a $((k,n))$ threshold scheme. A classic example is the problem of secure access to a bank vault. A bank manager might wish to require at least $k$ of their executives to jointly agree to open the vault using their independent keys. Unlike in classical protocols, the security of QSS is provided by the no-cloning theorem, which offers information-theoretic protection against unauthorized access to the secret state. One should differentiate between QSS protocols that share a classical secret with security guaranteed by the laws of quantum physics, which can be done without the use of entanglement~\cite{Richter2021}, and those that share a quantum secret (quantum state), as in this work.

QSS was first proposed for the discrete-variable (DV) regime by Cleve et al.~\cite{Cleve1999}, and later extended to the CV regime by Tyc and Sanders in 2002~\cite{Tyc2002}. That CV-QSS scheme focused on the particular problem of sharing a secret coherent state $\ket{\alpha}$ between $n$ players, of which at least $k$ must collaborate to reconstruct the secret. The protocol protects the secret information against internal and external adversaries acting alone~\cite{Tyc2002}. Early experiments in the optical regime have demonstrated the feasibility of CV-QSS using two-mode squeezed (TMS) resource states to share a secret coherent state $\ket{\alpha}$~\cite{Lance2004}. A generalized version of these protocols, allowing for the use of any generally-asymmetric Gaussian entanglement resource for the sharing of any arbitrary single-mode Gaussian state, has been recently devised in Ref.~\cite{Wilkinson2023}. This protocol is highly relevant to the development of robust and secure quantum communication and to the implementation of blind quantum computing (BQC).

%It has recently been shown that the quantum resource required for sharing CV quantum secrets is a strong form of entanglement termed quantum steering~\cite{Wilkinson2023}.

In this paper, we extend these ideas into the microwave frequency regime, $f \simeq 5$\,GHz. To emulate various elements of the QSS scheme, we construct a local tripartite microwave quantum network formed by superconducting coaxial cables connecting superconducting quantum circuits. We use superconducting Josephson parametric amplifiers (JPAs) for the generation of quantum-entangled resource states~\cite{Fedorov2018,Honasoge2025}. Using this entanglement as a resource, we implement a $((2,3))$ threshold QSS protocol with the measurement-device-free approach~\cite{Fedorov2021,Butt2025} and demonstrate reconstructed state fidelities exceeding the no-cloning threshold for a codebook of coherent states. The demonstrated results illustrate the growing maturity of superconducting quantum networks and their ability to exploit unique advantages provided by nonlinear superconducting circuits, such as JPAs, for various quantum communication tasks. 

The paper is structured as follows. First, in Sec.\,\ref{Section:Theory}, we provide a basic theoretical introduction to the principles of QSS and the information-theoretic concepts we will use to analyze our results. Then, in Sec.\,\ref{Section:ExperimentalSetup}, we describe our tripartite superconducting quantum network and its basic building blocks, such as JPAs and hybrid rings. Section\,\ref{Section:Performance} provides our central results demonstrating the effectiveness and security of microwave QSS. This includes a detailed analysis of experimental results in terms of communication security under various attack configurations. Finally, Secs.\,\ref{Section:DenseCoding} and \ref{Section:ErrorCorrection} demonstrate how the implemented QSS protocol can be reinterpreted for other valuable quantum information processing tasks, namely dense coding and erasure error correction, respectively. We conclude the paper with Sec.\,\ref{Section:Conclusion} by discussing the main results and outlining future perspectives.

\begin{figure*}[t!]
    \includegraphics[width=0.9\linewidth]{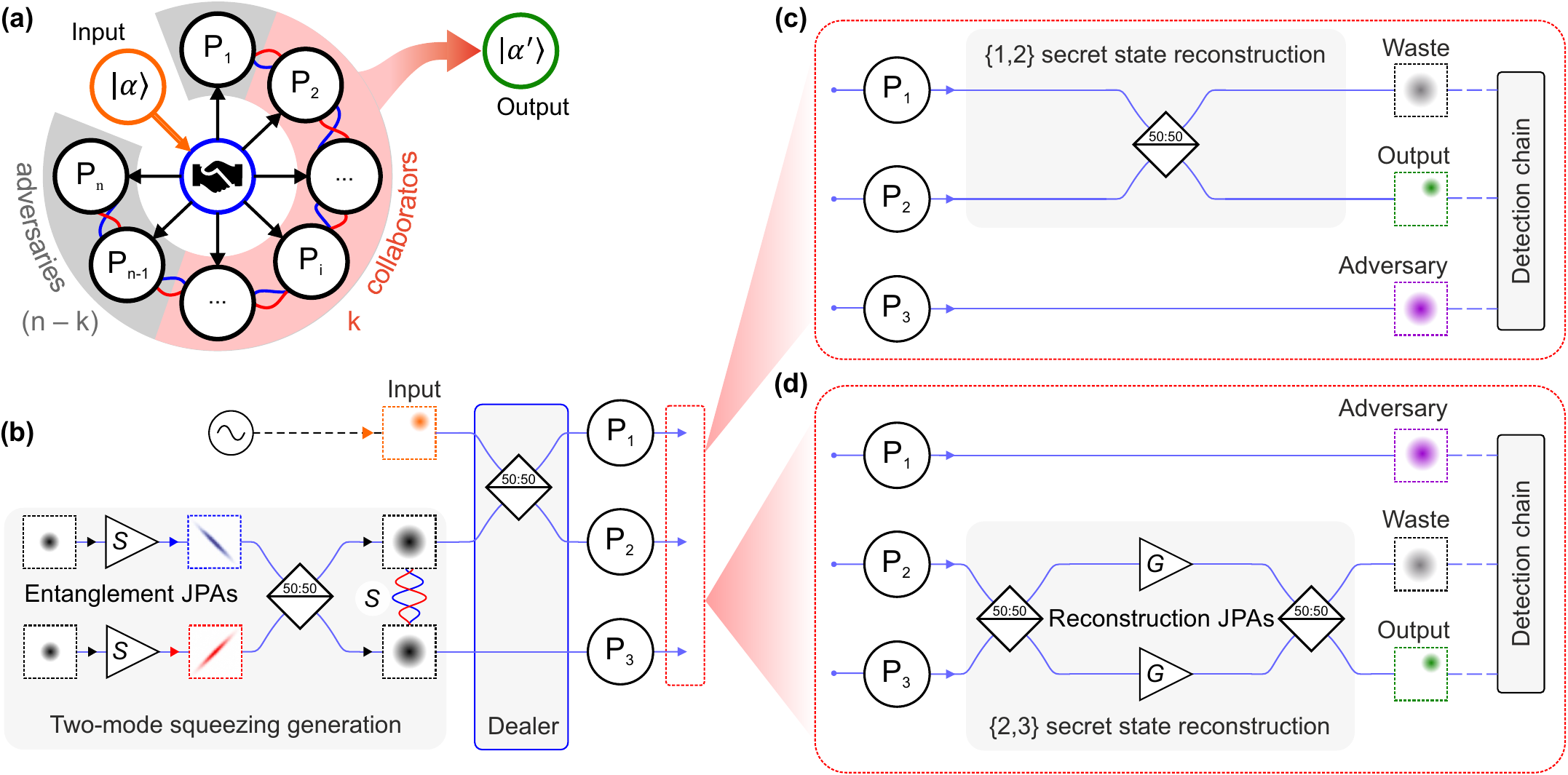}
    \caption{Schematic overview of the $((k,n))$ threshold QSS protocol implemented in this work. \textbf{(a)} Illustration of secret sharing among $n$ players, where a subset of $k$ players collaborate to securely reconstruct a secret input state $\ket{\alpha}$. \textbf{(b)} Experimental scheme of the Dealer part of the QSS protocol, where the Dealer superimposes an unknown secret coherent state with one mode of a TMS entanglement resource and distributes the outcome to the three players, $P_i$. The TMS entanglement resource is generated by two entanglement JPAs that produce orthogonally squeezed states with squeezing level $S$. The symmetric beam splitter operations are implemented using microwave hybrid rings. The input and resource states are depicted by their phase-space diagrams, showing respective Gaussian-distributed quantum uncertainties. Experimental schemes of two particular secret state reconstruction schemes investigated in our experiment: \textbf{(c)} the trivial $\{1,2\}$ scheme and \textbf{(d)} the nontrivial $\{2,3\}$ scheme. In the $\{2,3\}$ scheme, the input state is reconstructed using two reconstruction JPAs operated at degenerate gain $G$.}
    \label{Fig1}
\end{figure*}

\section{Theoretical overview} \label{Section:Theory}

\subsection{$((2,3))$ threshold QSS protocol}

A QSS scheme consists of two parts: a \emph{Dealer} protocol and a number of \emph{reconstruction} protocols, enabled by an entanglement resource (see Fig.\,\ref{Fig1}). The Dealer takes as input the original secret state, whose properties are unknown even to the Dealer, and an entangled resource state. Then, the Dealer produces a number of shares by mixing the original secret state with the resource [see Fig.\,\ref{Fig1}(b)]. Each of these shares is distributed to a single player, $P_i$. These shares are prepared in such a way that the original secret is individually obscured but can be retrieved through collaboration among a sufficient number of players.

%One of the primary aims of this paper will be to demonstrate that the collaborating players are able to reconstruct the secret state with greater accuracy than the adversaries.

Any sufficiently large number of players can then work together to reproduce the original state through a reconstruction scheme. Figures.\,\ref{Fig1}(c) and (d) show two such reconstruction schemes. We term the players in this authorized subgroup the ``collaborating players" and, assuming a worst-case scenario, all other players the ``adversaries".

We demonstrate in this paper a scheme termed $((2,3))$ threshold QSS, in which the secret state is split into three parts and can be reconstructed from any two of them. The Dealer mixes the original secret state, $\ket{\alpha}$, denoted as ``Input" in Fig.\,\ref{Fig1}(b), with one mode of the resource state via a balanced beam splitter (also known as a hybrid ring in the microwave implementation). The beam splitter operation results in three output modes, $\mode_i$, described by their respective bosonic annihilation operators as
\begin{align}
    \mode_1 &= \frac{1}{\sqrt{2}} \bigl( \mode_\alpha + \mode_\text{r1}), \\
    \mode_2 &= \frac{1}{\sqrt{2}} \bigl( \mode_\alpha - \mode_\text{r1}), \\
    \mode_3 &= \mode_\text{r2},
\end{align}
where $\mode_\alpha$ represents the input mode and $\mode_{\text{r}i}$ the $i$-th mode of the TMS resource state. In each individual output mode, the large quantum uncertainty from the resource state component obscures the properties of the secret state. The secret input information is thus hidden during transmission. Any two of these three modes can be used to reconstruct an approximation of the original secret state, with the reconstruction scheme dependent on which two shares are chosen. Should the owners of modes $\mode_1$ and $\mode_2$ collaborate (denoted as $\{1,2\}$ reconstruction), the operation is trivial: they can simply run the Dealer protocol in reverse to bring the system to its original state, retrieving the secret state as one of the outputs, as shown in Fig.\,\ref{Fig1}(c). Here, we term the $\{i,j\}$ reconstruction scheme to denote the procedure employed by collaborating players $P_i$ and $P_j$.

Reconstructing the secret with modes $\mode_1$ and $\mode_3$ ($\{1,3\}$ reconstruction) or modes $\mode_2$ and $\mode_3$ ($\{2,3\}$ reconstruction) requires a more complex setup. Here, the entanglement between the two resource modes is exploited to disentangle the $\mode_{\text{r1}}$ component from $\mode_\alpha$ in $\mode_1$ (or $\mode_2$, respectively). This could be physically implemented in a number of ways~\cite{Lance2005}, all of which ultimately construct the output state~\cite{Wilkinson2023} 
\begin{equation}
    \mode_{\rm out} = \eta\left( \sqrt{2} \mode_1 - \gamma \mode^\dagger_3 \right) = \eta \left( \mode_\alpha + \mode_\text{r1} - \gamma \mode^\dagger_\text{r2} \right)
\end{equation}
or
\begin{equation}
    \mode_{\rm out} = \eta\left( \sqrt{2} \mode_2 + \gamma \mode^\dagger_3 \right) = \eta \left( \mode_\alpha - \mode_\text{r1} + \gamma \mode^\dagger_\text{r2} \right),
\end{equation}
where $\eta = 1/\sqrt{2-\gamma^2}$ represents an amplification of the output state and $\gamma \in [ 0,\sqrt{2} )$ is selected to maximize reconstruction fidelity. Since the resource state is entangled such that $(\mode_\text{r1} - \gamma \mode^\dagger_\text{r2})$ overlaps destructively, the previously large variance is minimized. The size of this overlap is determined by the degree to which the resource state exhibits a strong form of entanglement termed EPR steering~\cite{Reid1988}. The quality of the reconstruction is therefore directly dependent on the quality of the entanglement resource state used.

Since the shares possessed by players 1 and 2 are symmetric [see $P_1$ and $P_2$ in Fig.\,\ref{Fig1}(b)], the $\{1,3\}$ and $\{2,3\}$ reconstruction schemes are also symmetric. Therefore, we can cover all classes by demonstrating only the $\{1,2\}$ and $\{2,3\}$ schemes. It has also been shown that the $\{1,2\}$ reconstruction scheme always outperforms the $\{1,3\}$ and $\{2,3\}$ schemes~\cite{Wilkinson2023}. As we are primarily interested in the worst-case performance of this protocol, we will focus on the $\{2,3\}$ scheme.

%\subsection{Gaussian codebook}

For the communication protocol, we will consider the transmission of coherent-state ensembles, where individual coherent amplitudes are drawn from a Gaussian distribution with mean zero and variance $\sigma^2$, hence forming a Gaussian codebook. Coherent states, $|\alpha \rangle$, are the eigenstates of the annihilation operator, $\hat a \vert \alpha \rangle = \alpha \vert \alpha \rangle$, with coherent amplitude $\alpha$. These states are described by Gaussian phase-space quasi-probability distributions and are consequently characterized entirely by their mean amplitude and second-order variances.

\subsection{Mutual information: security for trusted communication channels}

A QSS protocol is considered secure when the collaborating players are certain that they have the best possible copy of the original state. If we assume that the adversary has access only to their share --- no other information --- we can certify this security by directly comparing the information available at each output mode. This is equivalent to assuming that we can trust all quantum channels used by the two collaborating players.

The information available from a given output mode can be quantified by the quantum mutual information (MI) between the output state and the original input state~\cite{Braunstein2005,Serafini2017}. When the input coherent states are drawn according to a zero-mean Gaussian probability distribution with variance $\sigma^2$, the MI between the input and output modes can be simply written as~\cite{Yam2025}
\begin{equation}
    \textrm{MI} = \ln\left( 1 + \frac{4\sigma^2}{1+n_\textrm{eff}} \right),
\label{Eq:MI}
\end{equation}
where $\ln$ is the natural logarithm and $n_{\textrm{eff}}$ is the effective added noise in the reconstruction procedure (specified later in Sec.\,\ref{Subsection:AdversaryAttack}). For a known input state distribution, described by a known $\sigma^2$, the MIs obtained by the collaborating players, $\textrm{MI}_\textrm{collab}$, and the adversary, $\textrm{MI}_\textrm{adv}$, can be directly compared. The QSS protocol is considered secure when
\begin{equation}
    {\textrm{MI}_\textrm{collab} > \textrm{MI}_\textrm{adv}}, 
\label{MI-criterion}
\end{equation}
assuming that the adversary has access only to the information contained within their share.

\subsection{Fidelity: quality of QSS output and security for untrusted channels}

The condition in Eq.\,({\ref{MI-criterion}}) is a sufficient guarantee of security against opportunistic adversaries with access only to their allocated share. A more prepared attacker may also intercept portions of the other shares in a way that is functionally indistinguishable from transmission loss. To guarantee unconditional security against the full range of attacks, we further consider security grounded in the fundamental limits on the information that may exist about the secret state. We do this through the no-cloning theorem~\cite{Scarani2005,Wilkinson2023}, which provides the threshold at which the information contained in one mode is so large that any other mode must contain less information.

We will quantify the information content through the fidelity between the reconstructed output state and the original input state. A generalization of the Hilbert-Schmidt overlap, fidelity is a measure of the closeness of two quantum states based on how much their density matrices overlap, ranging between $F=0$ for orthogonal states to $F=1$ for fully identical states. For two Gaussian states, fidelity can be found as~\cite{Weedbrook2012,Scutaru1998}
\begin{equation}
	F(\alpha_1, V_1, \alpha_2, V_2) = \frac{1}{2} \frac{\exp\left[ -\frac{1}{2} d^\intercal (V_1+V_2)^{-1} d \right]}{\sqrt{\Lambda + \Delta} - \sqrt{\Delta}},
\label{Eq:fidelity}
\end{equation}
with
\begin{align}
	\Lambda &= \det(V_1+V_2), \\
	\Delta &= 16(\det V_1 - 1/16) (\det V_2 - 1/16), \\
	d &= \alpha_1 - \alpha_2,
\end{align}
where $\alpha_1$, $V_1$ and $\alpha_2$, $V_2$ are the complex displacement vectors and covariance matrices of states 1 and 2, respectively. In this paper, we use the vacuum variance definition of $1/4$ per field quadrature. Equation\,(\ref{Eq:fidelity}) quantifies how close the state reconstructed in our experiment is to the original quantum secret. We will use this fidelity to assess the performance of our protocol in Sec.\,\ref{Section:Performance}.

Combining this fidelity measure with the no-cloning theorem will allow us to remove the assumption that the communication channels are trusted, and thereby prove security more generally. It is well known that an unknown quantum state cannot be perfectly cloned. It has been shown, though, that such a state can be imperfectly cloned with a certain maximum fidelity termed the no-cloning threshold~\cite{Grosshans2001,Scarani2005}. To increase the fidelity of one clone above this threshold would require an equivalent reduction in the fidelity of another clone. In our protocol, should the authorized output state exceed this fidelity threshold, it is therefore guaranteed that any copy held by the adversary must contain less information --- regardless of how many sources of information the adversary has access to. This is a significantly stronger guarantee of security than that obtained by comparing the MIs, and exceeding the no-cloning threshold immediately certifies that the protocol is unconditionally secure.

For coherent states randomly drawn from the full spectrum of possible amplitudes, it has been shown that the no-cloning threshold is $F_\textrm{nc} = 2/3$~\cite{Grosshans2001}. As this represents the ability to clone a coherent state in the limit of infinite codebook size, we term this the asymptotic no-cloning fidelity. When coherent states are drawn from a finite codebook of possible states, the increase in classically-available information about the state distribution leads to a different achievable cloning fidelity~\cite{Cochrane2004,Yam2025}. For states drawn from a Gaussian codebook with variance $\sigma^2$, the no-cloning fidelity threshold is given by~\cite{Cochrane2004}
\begin{equation}
    F_\textrm{nc}(\sigma) =
    \begin{cases}
        \frac{4\sigma^2 + 2}{6\sigma^2 + 1} \qquad & \sigma^2 \ge \frac{1}{2} + \frac{1}{\sqrt{2}}; \\
        \frac{1}{(3 - 2\sqrt{2})\sigma^2 + 1} \qquad & \sigma^2 \leq \frac{1}{2} + \frac{1}{\sqrt{2}}.
    \end{cases}
\label{Eq:GaussianCodebook}
\end{equation}
When the states are drawn from a tighter distribution, such that $\sigma^2$ is smaller, it becomes easier to guess the input quantum state, so the corresponding no-cloning threshold is higher. In this paper, we analyze both the QSS protocol's ability to exceed the asymptotic threshold and the codebook-specific threshold.

\section{Experimental setup} \label{Section:ExperimentalSetup}

We implement the three-player QSS protocol, as introduced in Sec.\,\ref{Section:Theory}, using propagating microwave states with carrier frequency of $\SI{5.4}{\giga\hertz}$ inside a superconducting circuit setup, as schematically shown in Fig.\,\ref{Fig1}. The input secret states of this protocol are coherent states, defined by their complex displacement amplitudes $\alpha$. These coherent states are generated from room-temperature microwave sources, and then attenuated within the cryogenic setup to reach the single-photon level. Due to the finite background temperature of around $\SI{50}{\milli\kelvin}$, the input state has a weak thermal contribution. The entanglement resource is a TMS state produced by superimposing two orthogonally squeezed states with squeezing levels $S$ via a microwave hybrid ring~\cite{Fedorov2018,Zhong2013}. These individual squeezed states are generated using ``entanglement JPAs", which perform the two-mode squeezing operation when operated in the degenerate gain regime~\cite{Fedorov2018,Honasoge2025}. The Dealer step is implemented by mixing the input coherent state and one mode of the TMS state via another hybrid ring. This results in three player modes after the Dealer step, as shown in Fig.\,\ref{Fig1}(b), where modes $P_1$ and $P_2$ are symmetric.

For the $\{1,2\}$ reconstruction scheme, illustrated by Fig.\,\ref{Fig1}(c), the collaborating players $P_1$ and $P_2$ can undo the Dealer's beam splitter operation by using a second hybrid ring with matching phase relations. This is a trivial operation that, in the absence of losses and noise, would perfectly reconstruct the input state. In this scheme, the adversary player $P_3$ receives the TMS mode that has not interacted with the input state and hence contains no information about the secret.

For the $\{2,3\}$ reconstruction scheme, illustrated by Fig.\,\ref{Fig1}(d), the collaborating players $P_2$ and $P_3$ perform a two-mode squeezing operation on their modes. The two-mode squeezing operation is implemented using two ``reconstruction JPAs", both operated at a degenerate gain $G$, placed in a Josephson interferometer configuration~\cite{Pogorzalek2017,Kronowetter2023}. This reconstruction scheme utilizes destructive interference between the entangled modes of the TMS state to estimate the input state $\alpha$. Perfect reconstruction is only possible with infinite resource squeezing, but we will show that good reconstruction fidelity is achievable using realistic squeezing levels. In this scheme, the adversary $P_1$ receives a mode in which information about the secret state is obscured by the TMS noise. The $\{1,3\}$ reconstruction scheme may be implemented in the same way as the $\{2,3\}$ scheme because they are symmetric.

Thus, we can understand this QSS protocol as an encoding and decoding procedure. The Dealer enciphers the secret information using nonlocal correlations in the TMS resource state. In the $\{1,2\}$ scheme, the collaborating players decipher the secret by undoing the effect of the TMS correlations via a hybrid ring. In the $\{2,3\}$ scheme, the encoded secret is deciphered by removing the TMS noise via destructive interference with the other TMS mode. With only one of the three modes, the adversary cannot decode the secret information, so the protocol guarantees security.

In our experiment, the output states are characterized using quantum state tomography based on the measured field amplitudes~\cite{Menzel2010}. For this, we amplify the outgoing states, down-converting the $\SI{5.4}{\giga\hertz}$ signals to an intermediate frequency of $\SI{11}{\mega\hertz}$, and detect them using a field-programmable gate array receiver. The receiver has a measurement bandwidth of $\SI{400}{\kilo\hertz}$ and calculates statistical moments of the received signals up to the fourth order. We use the first- and second-order moments to determine the quantum states in the Gaussian approximation and then verify Gaussianity using higher-order moments. We calibrate the gain and noise of our detection chain using the Planck spectroscopy~\cite{Mariantoni2010,Gandorfer2025}. For each set of experimental operating parameters, we take on the order of $\SI{e8}{}$ averages to accurately estimate the quantum states.

\begin{figure*}[t!]
    \includegraphics[width=\linewidth]{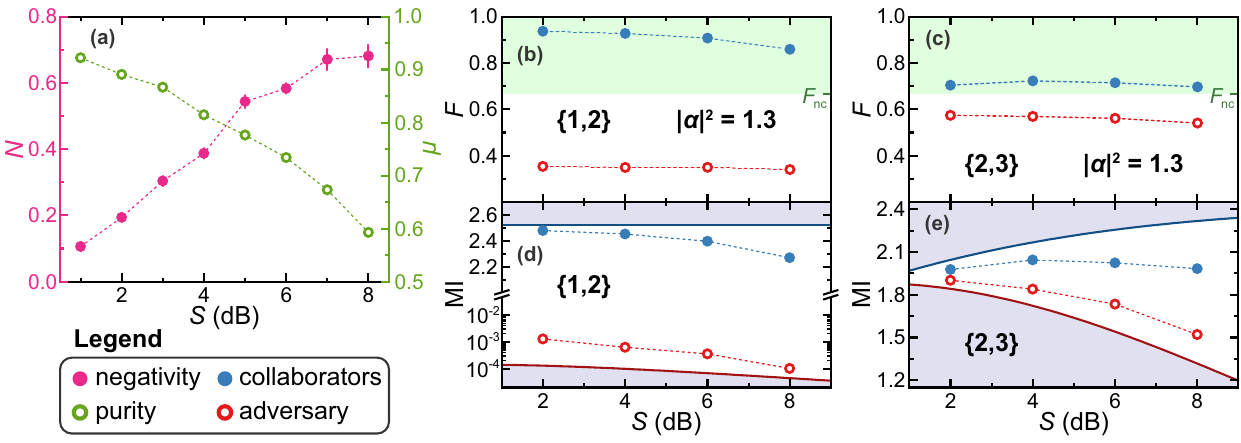}
    \caption{\textbf{(a)} Negativity $N$ and purity $\mu$ of the resource TMS state as a function of squeezing level $S$. Reconstructed state fidelities for the collaborating players and the adversary as a function of TMS resource squeezing level $S$, using \textbf{(b)} the $\{1,2\}$ scheme and \textbf{(c)} the $\{2,3\}$ scheme with reconstruction JPA gain $G = \SI{8}{dB}$. Input coherent states have average displacement photon number $|\alpha|^2 = 1.3$. Mutual information values obtained by the collaborating players and the adversary as a function of TMS resource squeezing level $S$, using \textbf{(d)} the $\{1,2\}$ scheme and \textbf{(e)} the $\{2,3\}$ scheme with reconstruction JPA gain $G = \SI{8}{dB}$. These MI values correspond to the communication of coherent states drawn from a Gaussian codebook with modulation variance $\sigma^2 = 3$. Blue solid lines represent the maximum MI obtainable by the collaborating players. Red solid lines represent the minimum MI accessible to the adversary. Error bars denote the standard error of the experimental data and are smaller than the symbol size when not shown.}
    \label{Fig2:FidelityInformation}
\end{figure*}

\section{QSS protocol performance} \label{Section:Performance}

We use the notion of fidelity introduced in Sec.\,\ref{Section:Theory} to assess the performance of the QSS state reconstruction. First, we characterize the quality of our TMS entanglement resource, as shown in Fig.\,\ref{Fig2:FidelityInformation}(a). We vary the squeezing level $S$ and observe the corresponding negativity $N$ and purity $\mu$ of the resulting TMS state. Negativity is an entanglement monotone derived from the Peres–Horodecki criterion, where $N>0$ implies entanglement~\cite{Peres1996,Adesso2005}, and purity describes the amount of classical noise admixed into a quantum state, where $\mu = 1$ implies a pure state. To achieve a high fidelity in QSS state reconstruction, we want a TMS resource with high $N$ and $\mu$. However, higher-order nonlinearities in our JPAs introduce additional noise (thereby a degradation in purities) at high squeezing levels~\cite{Boutin2017,Renger2021}, so we must seek a balance between sufficiently high entanglement and low noise. For our experiment, the best performance is reached at squeezing levels between $\SI{4}{}$ to $\SI{6}{\decibel}$.

Figures\,\ref{Fig2:FidelityInformation}(b) and \ref{Fig2:FidelityInformation}(c) show the reconstructed state fidelities for the $\{1,2\}$ and $\{2,3\}$ schemes, respectively. Here, we determine the state fidelities achieved by the collaborating players at various levels of resource squeezing, $S$. We also measure the fidelities that the adversary can reach if they use only their received share to reconstruct the secret state. We see that for input coherent states with $|\alpha|^2 = 1.3$, the collaborating players in both $\{1,2\}$ and $\{2,3\}$ schemes produce output states with fidelities above the asymptotic no-cloning threshold $F_\textrm{nc} = 2/3$~\cite{Grosshans2001}. Correspondingly, the adversary always receives states that have fidelities below $F_\textrm{nc}$, in the asymptotic limit of infinitely large codebooks. As expected, for the collaborating players, the output-state fidelity is higher for the $\{1,2\}$ reconstruction scheme than the $\{2,3\}$ scheme. Here, we choose the reconstruction JPA gain of $G = \SI{8}{\decibel}$ because it is near the theoretical optimum value, as explained in Sec.\,\ref{Subsection:NoCloning}.

In theory, the output state fidelities for the $\{1,2\}$ scheme should be independent of the entanglement resource properties, since it is possible to perfectly disentangle the TMS noise via a suitable beam splitter operation (performed with a hybrid ring). However, we observe that reconstruction fidelities decrease due to a small phase mismatch between the two hybrid rings in the experimental setup. For the $\{2,3\}$ scheme, output state fidelities initially increase as more resource squeezing is provided, with consequently better TMS interference. However, beyond a certain point, a continued increase of resource squeezing results in lower reconstruction fidelities due to the additional JPA noise at higher squeezing levels~\cite{Boutin2017,Renger2021}.

Fidelities achievable by the adversary are significantly lower than those of the collaborating players. In the $\{1,2\}$ reconstruction scheme, the adversary receives a state with near-zero displacement because it has not interacted with the input coherent state. In the $\{2,3\}$ scheme, the adversary receives a state that retains a mean-displacement equivalent to around half of that of the input coherent state, but with significant TMS noise. For both these scenarios, the adversary does not obtain fidelities above $F_\textrm{nc}$. Furthermore, due to the added TMS noise, which the adversary cannot eliminate, the mean-displacement is obscured, and very little information about the secret coherent state is obtainable.

%CW: Could we expand that to the average performance across the three possible reconstructions? This could then be compared very favourably to teleportation -- WY: Indeed QSS would achieve better state transfer fidelities compared to teleportation, but teleportation also has the added benefit that one communication channel is classical (hence robust against losses and noise) so the comparison might not be completely equivalent.

\subsection{Security: adversary mode attack} \label{Subsection:AdversaryAttack}

We consider the QSS protocol secure when the collaborating players have access to more information than the adversary. We can quantify this security by comparing the MI between their corresponding output mode and the input secret mode. For this, we utilize Eq.\,(\ref{Eq:MI}), where the effective added noise $n_{\textrm{eff}}$ is calculated by rescaling the measured quadrature noise of the outgoing states by the amplification (or attenuation) of the state reconstruction procedure. This is equivalent to the amount of added noise referenced at the input state and is given by
\begin{align}
    n_{\textrm{eff}} = 4 (v_{\textrm{out}} - v_{\textrm{in}}) \left| \frac{\alpha_{\textrm{in}}}{\alpha_{\textrm{out}}} \right|^2,
\end{align}
where $\alpha_{\textrm{in}}$, $v_{\text{in}}$ and $\alpha_{\textrm{out}}$, $v_{\text{out}}$ are the complex displacement amplitudes and noise variances of the input and output states, respectively.

For an arbitrary codebook of input coherent states, if the MI obtained by the collaborating players exceeds that of the adversary, then the QSS protocol is secure. While different codebooks may affect the nominal amount of MI, they do not change its ordinality. That is, their choice does not change whether the collaborating players or the adversary get more information. Figures\,\ref{Fig2:FidelityInformation}(d) and \ref{Fig2:FidelityInformation}(e) show the MI obtained by the collaborating players and the adversary for the $\{1,2\}$ and $\{2,3\}$ reconstruction schemes, respectively. Here, we consider a Gaussian codebook with $\sigma^2 = 3$, which is the modulation variance of the average coherent photon number. The codebook variance of $\sigma^2 = 3$ is chosen because of experimental constraints related to power saturation of the reconstruction JPAs~\cite{Yam2025}.

%CW:is (..) correct? or for experimental physicists maybe an indication of what the spread in coherent photon numbers would look like?}. -- WY:this would be the variance of average photon number in ``shot noise units"; I guess it is the same thing?

We observe that the collaborating players achieve higher MI than the adversary at all levels of resource squeezing, demonstrating the security of our QSS protocol against an adversary mode attack. We also model the idealized QSS protocol, as shown by solid lines in Figs.\,\ref{Fig2:FidelityInformation}(d) and \ref{Fig2:FidelityInformation}(e), where the only contributions to the effective noise $n_\textrm{eff}$ are the temperature fluctuations of the input state and the TMS entanglement resource. As seen in Figs.\,\ref{Fig2:FidelityInformation}(d) and \ref{Fig2:FidelityInformation}(e), our measured MI values fall within the regions bounded by the maximum MI obtainable by the collaborating players and the minimum MI accessible to the adversary.

%CW: Some closing comment on what this means is needed here. Also, we can very clearly say 'for the remainder of this paper we stop %considering the $\{1,2\}$ case'.

\begin{figure*}[t!]
    \includegraphics[width=\linewidth]{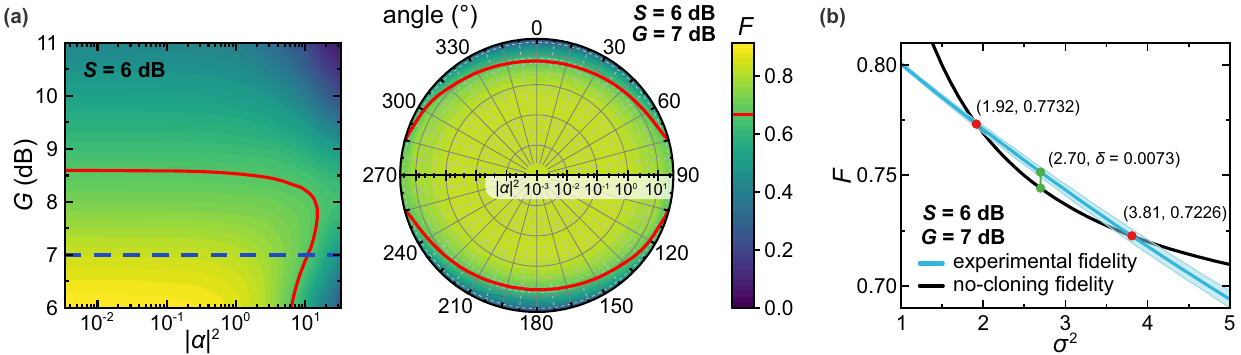}
    \caption{Security of the QSS implementation against an attack on all modes. \textbf{(a)} Reconstructed state fidelities for the $\{2,3\}$ scheme at various input coherent state displacements $|\alpha|^2$ and reconstruction JPA gains $G$. Fidelity distribution over the entire phase-space is shown for the operating parameters $S=\SI{6}{\decibel}$ and $G=\SI{7}{\decibel}$. Red solid lines indicate the asymptotic no-cloning threshold $F_\textrm{nc}=2/3$. \textbf{(b)} Average experimental fidelity compared to no-cloning fidelity for coherent states drawn from Gaussian codebooks with different variances $\sigma^2$. Red dots bound the parameter regime where genuine unconditional security is achieved. Green dots indicate the $\sigma^2$ where maximum security is obtained, characterized by an excess fidelity of $\delta = \SI{0.0073}{}$ above the codebook-specific no-cloning threshold. The blue-shaded region denotes the error bar of the experimental fidelity.}
    \label{Fig3:NoCloning}
\end{figure*}
% Solid lines represent theory model fits to the experimental QSS data. Error bars denote the standard error of the experimental data and are smaller than the symbol size when not shown.

\subsection{Security: attack on all modes} \label{Subsection:NoCloning}

For the aforementioned adversary mode attack, we determine the security of the QSS protocol in terms of the information obtained by the collaborating players and the adversary. However, to achieve unconditional security, we also need to consider attacks on all components of the protocol, not just the outgoing adversary mode. To this end, we utilize the no-cloning fidelity, derived from the no-cloning theorem~\cite{Grosshans2001}. If the collaborating players exceed the no-cloning fidelity with their reconstructed output state, they are guaranteed to have the best copy of the input secret state, and thus achieve unconditional security against any attack. Here, we further analyze only the $\{2,3\}$ reconstruction scheme, as depicted in Fig.\,\ref{Fig1}(d), since the $\{1,2\}$ scheme is the more trivial case and generally gives higher fidelities than the $\{2,3\}$ scheme. 

Figure\,\ref{Fig3:NoCloning}(a) shows the output state fidelities, at the collaborating players, for various input state displacements $|\alpha|^2$ and reconstruction gains $G$. When operating the reconstruction JPAs at $G = \SI{6}{}$ to $\SI{8}{\decibel}$, we obtain reconstruction fidelities above $F_\textrm{nc} = 2/3$ up to around $|\alpha|^2 = \SI{10}{}$. While the theoretical optimum value is $G = \frac{\sqrt{2}+1}{\sqrt{2}-1} \approx \SI{7.6}{\decibel}$~\cite{Wilkinson2023}, we observe that it is advantageous to use smaller gain values when reconstructing states with low $|\alpha|^2$. This is because operating the JPAs at lower gain introduces less JPA noise in the reconstruction step, at the cost of a slight mismatch in displacement. We can closely fit the experimental data with our theory model based on the beam splitter formalism. From this model, a simple expression can be obtained for the output state fidelity
\begin{equation}\label{Eq:FidelityExpression}
    F(\alpha,k,v_\textrm{out}) = \frac{2}{1 + 4v_\textrm{out}} \exp\left[ -2 \frac{(\sqrt{k}-1)^2}{1 + 4v_\textrm{out}} |\alpha|^2 \right],
\end{equation}
where $k$ is the overall gain and $v_\textrm{out}$ the noise variance at the output state of the QSS procedure.

%, which gives the solid lines in Fig.\,\ref{Fig3:NoCloning}(a)
%CW: Let us now consider the specific setup optimised for a Gaussian codebook of width $\sigma^2\approx3$...

To genuinely achieve secure communication, we must average over the reconstruction fidelities of states drawn from an experimentally realizable codebook. We consider a Gaussian codebook, whose no-cloning fidelity is given by Eq.\,(\ref{Eq:GaussianCodebook}) and depends on the codebook variance $\sigma^2$. We choose the reconstruction gain $G = \SI{7}{\decibel}$, since it gives the best overall fidelity for a Gaussian codebook with $\sigma^2 \approx 3$. For an ideal communication protocol, the output-state fidelities should be independent of the coherent state phase. We see in Fig.\,\ref{Fig3:NoCloning}(a) that our implementation of the QSS protocol is largely symmetric in phase up to $|\alpha|^2 \approx \SI{10}{}$. The asymmetry at higher displacements is due to a slight imbalance in the Josephson interferometer used for state reconstruction~\cite{Kronowetter2023}, arising from different noise properties in the reconstruction JPAs. Figure\,\ref{Fig3:NoCloning}(b) shows that average reconstructed state fidelities exceed the no-cloning fidelity defined by Eq.\,(\ref{Eq:GaussianCodebook}), thus providing unconditional security with $F > F_\textrm{nc}(\sigma)$. Here, the average experimental fidelity is calculated by integrating with respect to $|\alpha|^2$ over the theory model fit to Fig.\,\ref{Fig3:NoCloning}(a), simultaneously weighted by a Gaussian codebook distribution with variance $\sigma^2$. In our experimental implementation, unconditional security can be achieved for Gaussian codebooks with variances between $1.92 \le \sigma^2 \le 3.81$ and reaches a maximum at $\sigma^2 = 2.70$, characterized by an excess fidelity of $\delta = \SI{0.0073}{}$ above the codebook-specific no-cloning threshold. Thus, we demonstrate that such a QSS protocol is not only secure against an adversary-mode attack but also against all possible attacks.
\begin{figure*}[t!]
    \includegraphics[width=\linewidth]{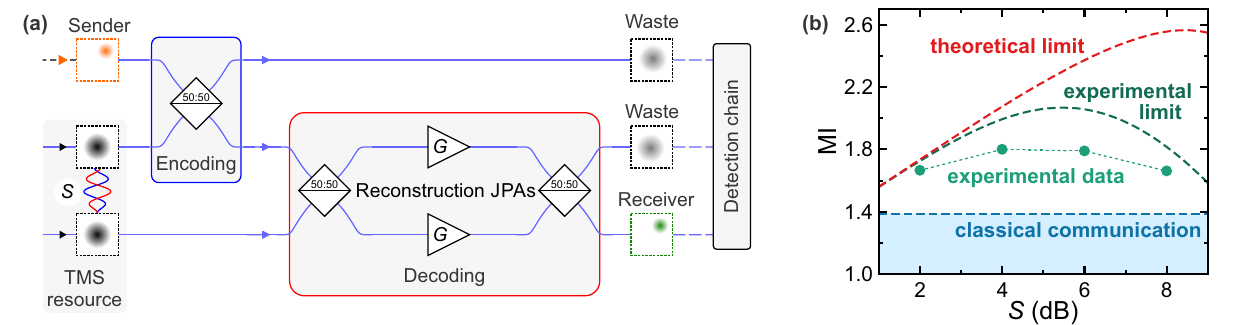}
    \caption{Dense coding as a reinterpretation of the $\{2,3\}$ reconstruction scheme in the QSS protocol. \textbf{(a)} Experimental scheme. The sender uses a hybrid ring to encode classical information from the input signal by locally displacing one part of the entangled TMS state. The encoded state is transmitted to the receiver, who uses the other part of the entangled TMS state to perform the decoding operation. By applying a suitable reconstruction gain $G$, the receiver can increase the signal-to-noise ratio beyond what is classically possible, thereby enhancing information transmission. \textbf{(b)} Measured MI values at the receiver as a function of the TMS squeezing, $S$, and at the reconstruction gain $G = \SI{8}{\decibel}$ and fixed Gaussian ensemble variance $\sigma_\textrm{ens}^2 = 3$. Green markers represent the measured MI values using dense coding. The blue-dashed line represents the MI corresponding to ideal single-mode coherent-state communication. The green-dashed line represents the MI limit for our experimental implementation of dense coding, taking into account the noise properties of our TMS resource state. The red-dashed line represents the theoretical maximum MI for the optimal implementation of dense coding. Error bars are smaller than the symbol size.}
    \label{Fig4:DenseCoding}
\end{figure*}

\section{Dense coding} \label{Section:DenseCoding}

The $\{2,3\}$ reconstruction scheme of QSS can also be adapted to work as a dense coding protocol, as schematically illustrated in Fig.\,\ref{Fig4:DenseCoding}(a). Here, a pre-distributed entanglement resource is used to enhance the amount of classical information that can be transmitted through a communication channel without changing the channel properties~\cite{Braunstein2000}. By consuming this entanglement resource via destructive interference, we can effectively squeeze the signal along both quadratures simultaneously. Thus, we can increase a corresponding signal-to-noise ratio and send more information per channel use than classically possible. For the particular case of CV dense coding, it is possible to send up to twice the information content as compared to single-mode, coherent state communication in the limit of high resource squeezing ~\cite{Braunstein2000,Mizuno2005}.

In dense coding, similar to the QSS procedure, the TMS resource state is first distributed between two modes. Then, one part of the TMS state is displaced by an external coherent input signal from the sender, which encodes classical information into the quantum state. In our implementation, this step is performed using a hybrid ring, as shown in Fig.\,\ref{Fig4:DenseCoding}(a). This state that provides the displacement amplitude corresponds to the secret coherent state in the QSS protocol. Now, we can interpret $P_2$ as the sender and $P_3$ as the receiver (see Fig.\,\ref{Fig1}). The information-carrying state at $P_2$ is then transmitted to the receiver, who can reconstruct the mean-displacement information via the two-mode squeezing operation of the $\{2,3\}$ scheme. By utilizing the pre-distributed TMS correlations in the state at $P_3$, the receiver can obtain more information than that which is transmitted solely through the state at $P_2$. Thus, we accomplish the task of dense coding.

To complete the analysis, we consider a communication channel that allows a Gaussian ensemble variance $\sigma_\textrm{ens}^2 = 3$. The ensemble variance characterizes the classical channel capacity and is given by~\cite{Braunstein2000}
\begin{equation}
    \sigma_\textrm{ens}^2 = \sigma_\textrm{cb}^2 + \sigma_\textrm{st}^2,
\end{equation}
where $\sigma_\textrm{cb}^2$ is the modulation variance for a Gaussian codebook and $\sigma_\textrm{st}^2$ is the noise variance of the transmitted states. In the ideal (noiseless and lossless) case, the state variance is just determined by the local TMS variance, $\sigma_\textrm{st}^2 = \sinh(2r)$ with $r = S/20 \ln(10)$ the squeezing parameter~\cite{Braunstein2000}. In an experiment, we can measure the output state variance, which includes additional noise contributions from the experimental setup. For a fair comparison between the dense coding protocol and direct single-mode communication, we must fix the ensemble variance $\sigma_\textrm{ens}^2$ for both scenarios. Since the state variance $\sigma_\textrm{st}^2$ increases as a function of TMS squeezing level in the dense coding implementation, this implies that codebook variance $\sigma_\textrm{cb}^2$ decreases correspondingly. For the direct single-mode communication, we consider the ideal case of transmitting perfect coherent states with $\sigma_\textrm{st}^2 = 0.25$. Realistic implementations of direct coherent state communication would result in larger $\sigma_\textrm{st}^2$ due to experimental losses and noise. Note that in our calculations for the QSS protocol in Sec.\,\ref{Section:Performance}, we maintain the same codebook modulation variance across all TMS squeezing levels, so the ensemble variances increase instead. This results in slightly different MI values calculated for these two protocols.

Figure\,\ref{Fig4:DenseCoding}(b) shows the performance of dense coding in our experimental implementation. As a baseline, we consider the ideal transmission (through a noiseless and lossless channel) of perfect coherent states, which is indicated by the blue-dashed line in Fig.\,\ref{Fig4:DenseCoding}(b). We analyze the achievable MI for our experimental implementation of dense coding, taking into account the noise properties of our TMS resource state [green-dashed line in Fig.\,\ref{Fig4:DenseCoding}(b)]. Moreover, we consider the theoretical limit of an ideal CV dense coding protocol with MI given by~\cite{Braunstein2000}
\begin{equation}
    \textrm{MI}_\textrm{dc} = \ln\left( 1 + \sigma_\textrm{cb}^2 e^{2r} \right),
\end{equation}
which is indicated for a fixed $\sigma_\textrm{ens}^2 = 3$ by the red-dashed line in Fig.\,\ref{Fig4:DenseCoding}(b). We see that a higher MI value can be achieved when using the entanglement resource, compared to the ideal coherent-state communication with the same ensemble variance $\sigma_\textrm{ens}^2$. This provides experimental evidence of quantum enhancement when using the QSS procedure for dense coding and illustrates the various capabilities of delocalizing information using TMS resource states.

\begin{figure*}[t!]
    \includegraphics[width=\linewidth]{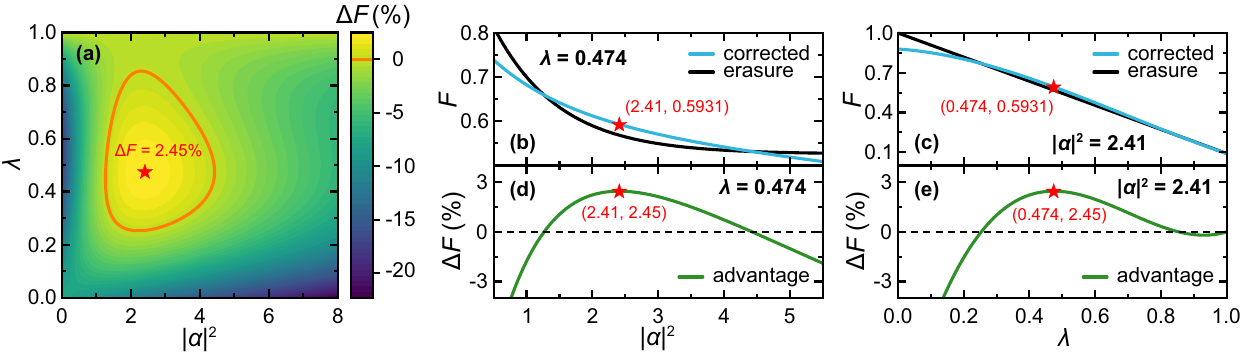}
    \caption{QSS for quantum error correction of erasure errors, operating at $S = \SI{6}{\decibel}$ and $G = \SI{7}{\decibel}$. \textbf{(a)} Fidelity advantage $\Delta F$ of using the QSS protocol for error correction as a function of input state displacement $|\alpha|^2$ and erasure probability $\lambda$. Solid orange line denotes the parameter regime where $\Delta F = \SI{0}{\percent}$. The maximum advantage of $\Delta F = \SI{2.45}{\percent}$ is achieved at $|\alpha|^2 = 2.41$ and $\lambda = 0.474$, as indicated by the red star. Fidelities as a function of \textbf{(b)} input state displacement and \textbf{(c)} erasure probability. Black and blue solid lines represent the ideal single-mode erasure channel and the simulated error correction result, respectively. Fidelity advantage as a function of \textbf{(d)} input state displacement and \textbf{(e)} erasure probability. Error bars are smaller than the line width.}
    \label{Fig5:ErrorCorrection}
\end{figure*}

\section{Erasure error correction} \label{Section:ErrorCorrection}

Our implementation of the QSS protocol can also be used for quantum error correction of erasure errors~\cite{Niset2008,Lassen2010,Lu2016}. In this scenario, all players collaborate, but each has a probability of losing their share of the original state. We consider that each player, represented by a corresponding transmission path in the experiment, has an independent erasure probability $\lambda$. Here, erasure means that a transmitted state is completely lost and replaced by a vacuum state. While the $\SI{50}{\milli\kelvin}$ background temperature causes a weak thermal distribution, this contribution is very small and hence well-approximated by the vacuum state. The $((k,n))$ threshold QSS protocol is able to correct for such erasure errors because, when a subset of transmission channels experience erasures, the original input state can still be recovered if the number of remaining channels fulfills $k > n/2$. This process is enabled by the delocalization of information via quantum entanglement.

\begin{table}[t]
    \centering
    \caption{Combinations of channel erasures and their corresponding probabilities. For each combination, we denote the best QSS reconstruction scheme that yields the highest state fidelity. Note that the $\{1,3\}$ and $\{2,3\}$ schemes are symmetric and provide same fidelity values.}
    \vspace{1em}
    \setlength{\tabcolsep}{6pt}
    \renewcommand{\arraystretch}{1.4}
    \begin{tabular}{c|c|c}
        \hline\hline
        Erasures & Probability & Best reconstruction scheme \\
        \hline
        none & $(1-\lambda)^3$ & $\{1,2\}$ collaborators \\
        $P_1$ & $\lambda (1-\lambda)^2$ & $\{2,3\}$ collaborators \\
        $P_2$ & $\lambda (1-\lambda)^2$ & $\{1,3\}$ collaborators \\
        $P_3$ & $\lambda (1-\lambda)^2$ & $\{1,2\}$ collaborators \\
        $P_1, P_2$ & $\lambda^2 (1-\lambda)$ & $\{1,2\}$ adversary \\
        $P_1, P_3$ & $\lambda^2 (1-\lambda)$ & $\{1,3\}$ adversary \\
        $P_2, P_3$ & $\lambda^2 (1-\lambda)$ & $\{2,3\}$ adversary \\
        $P_1, P_2, P_3$ & $\lambda^3$ & vacuum state \\
        \hline\hline
    \end{tabular}
    \label{Tab1:erasureCombinations}
\end{table}
Using our QSS experimental data, we can simulate erasure errors by assigning an erasure probability $\lambda$ to each transmission path. Since there are three paths in our experimental implementation, corresponding to the three players in QSS, there are eight possible combinations of erasure errors with probabilities ranging from $(1-\lambda)^3$ (none of the channels are erased) to $\lambda^3$ (all channels are erased). These combinations of channel erasures are summarized in Table\,\ref{Tab1:erasureCombinations}. For each erasure combination, we choose a QSS reconstruction scheme that yields the highest fidelity based on our measured data. In the cases where two channels are erased, we can only reconstruct the original state based on a single remaining channel, corresponding to the adversary mode. When all three channels are erased, the resulting state is the vacuum state. This leads to the average transmission fidelity
\begin{equation}
	\bar{F}(\alpha,\lambda) = \sum_{i=1}^{8} p_i(\lambda) F_i(\alpha),
\label{Eq:Erasure}
\end{equation}
where $p_i(\lambda)$ is the probability of a certain combination of erasures and $F_i(\alpha)$ is the corresponding experimental fidelity. As a baseline, we consider the transmission of perfect coherent states through a single-mode channel that has erasure probability $\lambda$ but is otherwise noiseless and lossless. The corresponding transmission fidelity is given by
\begin{equation}
    F_\textrm{coh}(\alpha,\lambda) = 1 - \lambda (1 - e^{-|\alpha|^2}).
\end{equation}
Given the same erasure probability $\lambda$, if the error-corrected fidelity $\bar{F}(\alpha,\lambda)$ is greater than that obtainable from a single-mode channel $F_\textrm{coh}(\alpha,\lambda)$, then our scheme achieves the desired quantum error correction capability.

Figure\,\ref{Fig5:ErrorCorrection} shows the fidelity advantage $\Delta F(\alpha,\lambda) = \bar{F}(\alpha,\lambda) - F_\textrm{coh}(\alpha,\lambda)$ by using our QSS implementation for the correction of erasure errors. Here, we choose the operating parameters $S = \SI{6}{\decibel}$ and $G = \SI{7}{\decibel}$, because this gives the best overall fidelities. The parameter regime where QSS provides an advantage, $\Delta F(\alpha,\lambda) > 0$, is denoted by the orange line in Fig.\,\ref{Fig5:ErrorCorrection}(a). The baseline fidelity of ideal coherent states transmitted through an erasure channel is represented by the black lines in Figs.\,\ref{Fig5:ErrorCorrection}(b) and (c). We observe an improvement in transmission fidelities of up to $\SI{2.45}{\percent}$, as indicated by the red stars in Fig.\,\ref{Fig5:ErrorCorrection}. Interestingly, this improvement exists only in the intermediate range of input state displacements $|\alpha|^2$ and erasure probabilities $\lambda$. We explain this observation by noting that for small $|\alpha|^2$ and $\lambda$, the idealized baseline scenario allows high fidelities, while the measured fidelities are limited by noise and losses from experimental elements, such as JPAs or hybrid rings. In the regime of large $|\alpha|^2$ and $\lambda$, the measured QSS fidelities deteriorate more quickly due to slight asymmetries in the experimental setup and power saturation of the reconstruction JPAs. In these regimes, the performance can be improved by methods such as using superconducting parametric devices with better 1-dB compression points or by using superconducting hybrid rings. Nonetheless, even our suboptimal QSS experiment clearly demonstrates the ability to correct for erasure errors and contributes towards efforts of building fault-tolerant quantum networks.

\section{Conclusion and Outlook} \label{Section:Conclusion}

In this work, we experimentally demonstrate a $((2,3))$ threshold QSS protocol in a local tripartite superconducting network at microwave frequencies. Our particular implementation relies on superconducting JPAs for both the generation and manipulation of microwave entangled signals, which enables the reconstruction of secret quantum coherent states with fidelities beyond the no-cloning threshold. We perform a security analysis that covers two eavesdropping scenarios, namely a local adversary attack and a general attack on all communication channels. In both scenarios, we are able to achieve unconditionally secure communication with regard to the attack, characterized by ${\textrm{MI}_\textrm{collab} > \textrm{MI}_\textrm{adv}}$ and $F > F_\textrm{nc}(\sigma)$, respectively. 

Furthermore, we investigate inherent connections between the QSS protocol and other quantum information processing protocols, which are fundamental to a future quantum internet, such as quantum dense coding and quantum erasure correction. For dense coding, we utilize a pre-distributed entanglement resource to enhance the classical communication rate above the ideal limit obtainable with a classical channel alone. For erasure error correction, we exploit the ability of QSS to delocalize the initial quantum information among separate communication channels, thereby compensating for complete signal erasure in a subset of these channels. This error correction feature of QSS is particularly valuable when dealing with realistic lossy transmission channels in experimental quantum networks.

In a more general sense, the $((k,n))$ threshold QSS protocol is an extremely powerful and versatile building block for genuinely multipartite networks. This protocol provides unconditional security for communications, resilience against localized errors, and high-fidelity transmission of quantum states. These properties are especially relevant for applications such as conference key agreement~\cite{Liu2023}, distributed quantum sensing~\cite{Ho2025}, and blind quantum computing~\cite{Barz2012,Fitzsimons2017}.

Blind quantum computing represents a particularly interesting and important direction for developing superconducting quantum networks due to their affinity with scalable superconducting quantum computers. In BQC, quantum computation can be carried out so that the quantum states and computational processes remain private and unknown to the servers on which the computation runs. This approach is crucial in scenarios when a client does not trust the server. Connections between quantum blindness and quantum shared secrets were first explored in the DV context~\cite{Ouyang2017}. The main incentive for harnessing QSS for BQC is illustrated by the fact that input quantum states are not decipherable during transit (i.e., in the quantum channels of the individual players). Here, the most straightforward way to arrange BQC is to split the entire computation process into separate sequences of logical gates, each applied to the individual QSS shares rather than the original input state. In principle, the reconstructed state via QSS should correspond to an output state as if the quantum computation was performed directly on the input state, but the servers fundamentally cannot extract any information about these quantum states from their local observables. Furthermore, if a subset of the servers boycott the computation and do not return their shares, we may still obtain the correct output by using the resilience of our $((k,n))$ threshold QSS protocol to erasures. In order to accomplish universal quantum computation, at least one non-Clifford gate is required. This has already been analyzed for the DV regime, and it is theoretically shown that the $((n,n))$ threshold QSS scheme allows the evaluation of quantum circuits on a shared secret state without needing to decode the secret, given that a set of ``magic states'' is used to enact non-Clifford gates~\cite{Ouyang2017}. The implementation of non-Clifford gates in the CV regime and the investigation of an optimal BQC scheme are subjects of future work.

It is worthwhile to note that QSS can be performed on graph states, also known as cluster states, which has been demonstrated for DV photonic networks~\cite{Bell2014}. Recently, the first implementation of programmable microwave cluster states via Josephson meta-materials has been reported~\cite{Alocco2025}. This opens up a completely new avenue for complex microwave circuits.

The concept of multipartite microwave quantum networks is a central feature in our work. In this paper, we demonstrate for the first time the continuous-variable quantum secret sharing protocol in a microwave tripartite superconducting network. We have also shown how QSS already enables other fundamental quantum information processing tasks, such as quantum dense coding and quantum error correction. Ultimately, QSS has the potential to be extended towards the BQC paradigm, thus demonstrating the capability of QSS as a ``Swiss Army knife" for quantum networks.

\acknowledgments

We acknowledge support by the German Research Foundation via Germany`s Excellence Strategy (EXC-2111-390814868), the German Federal Ministry of Education and Research via the project QUARATE (Grant No.~13N15380), the Scottish Universities Physics Alliance (SUPA), and the Global Ph.D. program of the University of St.\,Andrews and Macquarie University. This research is part of the Munich Quantum Valley, which is supported by the Bavarian state government with funds from the Hightech Agenda Bayern Plus. We thank Jianyu Tang for assistance in analyzing potential BQC schemes.

\bibliography{Bibliography} 

\end{document}